# Chandra Observation of an X-ray Flare at Saturn: Evidence for Direct Solar Control on Saturn's Disk X-ray Emissions

Anil Bhardwaj[1,*], Ronald F. Elsner[1], J. Hunter Waite, Jr.[2], G. Randall Gladstone[3], Thomas E. Cravens[4], and Peter G. Ford[5]

[1] NASA Marshall Space Flight Center, NSSTC/XD12, 320 Sparkman Drive, Huntsville, AL 35805; anil.bhardwaj@msfc.nasa.gov, ron.elsner@msfc.nasa.gov
[2] Department of Atmospheric, Oceanic and Space Sciences, University of Michigan, Ann Arbor, MI 48109; hunterw@umich.edu
[3] Southwest Research Institute, San Antonio, P.O. Drawer 28510, TX 78228; randy.gladstone@swri.org
[4] Department of Physics and Astronomy, University of Kansas, Lawrence, KS 66045; cravens@ku.edu
[5] Massachusetts Institute of Technology, Kavli Institute for Astrophysics and Space Research, 70 Vassar Street, Cambridge, MA 02139; pgf@space.mit.edu

[*] on leave from: Space Physics Laboratory, Vikram Sarabhai Space Centre, Trivandrum 695022, India; bhardwaj_spl@yahoo.com

Running header: Saturn X-ray Emission

# ABSTRACT

Saturn was observed by Chandra ACIS-S on 20 and 26-27 January 2004 for one full Saturn rotation (10.7 hr) at each epoch. We report here the first observation of an X-ray flare from Saturn's non-auroral (low-latitude) disk, which is seen in direct response to an M6-class flare emanating from a sunspot that was clearly visible from both Saturn and Earth. Saturn's disk X-ray emissions are found to be variable on time scales of hours to weeks to months, and correlated with solar F10.7 cm flux. Unlike Jupiter, X-rays from Saturn's polar (auroral) region have characteristics similar to those from its disk. This report, combined with earlier studies, establishes that disk X-ray emissions of the giant planets Saturn and Jupiter are directly regulated by processes happening on the Sun. We suggest that these emissions could be monitored to study X-ray flaring from solar active regions when they are on the far side and not visible to Near-Earth space weather satellites.

*Subject heading*: planets and satellites: individual (Saturn) – X-rays: general – Sun: flares – scattering – Sun: X-rays

## 1. Introduction

Saturn and Jupiter are alike in many respects — both are giant magnetized planets with huge magnetospheres and similar rotation rates. However, Saturn's magnetosphere is thought to be powered in a fashion intermediate between the Jovian case, which is largely driven by the rapid rotation of Jupiter's huge magnetic field with large internal sources of plasma, and the Earth's magnetosphere, whose dynamics are dominated by solar wind interactions (Russell 2001; Waite & Lummerzheim 2002; Krupp et al. 2004; Clarke et al. 2005). Unlike Jupiter (offset angle 9.6°) or the Earth (11.5°), the magnetic dipole of Saturn is closely aligned with the planetary spin axis (<1°). Auroral emissions from Saturn have been observed at far ultraviolet (FUV), infrared and radio wavelengths (Bhardwaj & Gladstone 2000a, 2000b; Gérard et al. 2004; Stallard et al. 2004; Kurth et al. 2005; Clarke et al. 2005). Early attempts to detect X-ray emission from Saturn with Einstein in December 1979 (Gilman et al. 1986) and ROSAT in April 1992 (Ness & Schmitt 2000) were negative and marginal, respectively. Saturnian X-rays were unambiguously observed by XMM-Newton in October 2002 (Ness et al. 2004a) and by the Chandra X-ray Observatory (CXO) in April 2003 (Ness et al. 2004b).

Both Jupiter and Earth exhibit X-ray emission associated with their auroras and their non-auroral (low-latitude) disks (see reviews by Bhardwaj et al. 2002; Waite & Lummerzheim 2002; Petrinec et al. 2000). The X-ray aurora on Earth is mainly generated by electron bremsstrahlung (e.g., Stadsnes, Aarsnes, & Bjordal 1997; Petrinec et al. 2000; Bhardwaj et al. 2005b), while Earth's disk X-ray emissions are known to be produced due to scattering and fluorescence of solar X-rays (McKenzie, Rugge, & Charles 1982; Snowden & Freyberg 1993; Petrinec et al. 2000). Recent X-ray observations of Jupiter by Chandra and XMM-Newton have confirmed, as proposed earlier (for review, see Bhardwaj and Gladstone 2000a), that auroral X-rays of Jupiter are produced by charge-exchange of precipitating highly-ionized heavy ions (Branduardi-Raymont et al. 2004; Elsner et al. 2005). However, these ions are not from the inner (~8–12 $R_J$; $R_J$ = Jupiter radius) magnetosphere, as thought earlier, but are from the outer (>30 $R_J$) magnetosphere or/and solar wind, and apparently undergo large acceleration to attain energies of >1 MeV/nucleon before impacting on the Jovian upper atmosphere (Gladstone et al. 2002; Cravens et al. 2003; Elsner et al. 2005). The disk X-ray emissions from Jupiter were initially proposed to be largely due to the precipitation of ions from the radiation belts (Waite et al. 1997), but recent observations (Gladstone et al. 1998; Bhardwaj et al. 2004, 2005a; Branduardi-Raymont et al. 2004) suggest that they are resonant and fluorescent scattered solar X-rays (Maurellis et al. 2000; Bhardwaj et al. 2004, 2005a; Cravens et al. 2005).

In this letter, we present the first observation of an X-ray flare from Saturn's disk. We also demonstrate that, like on Jupiter, solar processes govern the production of X-ray emission from the disk of Saturn.

## 2. Observations

In January 2004 Saturn was observed by the Advanced CCD Imaging Spectrometer (ACIS) of the Chandra observatory in two exposures, 00:06 to 11:00 UT on 20 January, and 14:32 UT on 26 January to 01:13 UT on 27 January. ACIS was configured with the spectroscopic array in imaging mode and the planetary image was confined to Node B of the S3 CCD, the detector with the greatest sensitivity to X-ray energies below 1 keV. Pulse-height values for individual X-ray events were corrected for effects due to Saturn's optically bright disk (cf. Elsner et al. 2002); these corrections were much smaller than for the case of Jupiter (Elsner et al. 2005). Chandra events are time tagged and were therefore mapped into Saturn's rest frame using ephemerides for the dates above created using the online JPL HORIZONS ephemerides generator. The space at and around the planet was divided into regions covering the portion of the planet not obscured by the rings, the south polar cap region, and the unobscured planet minus the south polar cap (cf. Figure 1). In the energy range 0.24–2.0 keV, where essentially the entire planet's X-ray emission is detected, the number of events registered on 20 January (26 January) from unobscured disk (not obscured by rings and no south polar cap region) and south polar cap regions are 134 (32) and 17 (6), respectively. The expected background counts for these regions scaled from a large region outside the planet on 20 January (26 January) are 38.4 (2.7) and 5.2 (0.4), respectively. Since the planet blocks X-rays from beyond, leaving only background events induced by charged particles, these background numbers are upper limits to the true background. With a total integration time of 37.37 (36.3) ks on 20 January (26 January), respectively, the net count rate for these regions are $2.56\pm0.31 \times 10^{-3}$ ($3.84\pm1.1 \times 10^{-4}$) and $7.37\pm1.6 \times 10^{-4}$ ($1.55\pm0.67 \times 10^{-4}$) cts/s, respectively. During the observation, Saturn's angular diameter and its heliocentric and geocentric distances were 20.5″, 9.03 AU and 8.11 AU, respectively, on January 20, and 20.4″, 9.03 AU and 8.15 AU, respectively, on January 26.

## 3. Results and Discussion

Figure 1 shows the distribution of X-ray brightness over Saturn during the January 20 and 26, 2004 Chandra ACIS-S observation in the energy range 0.24–2.0 keV. These observations suggest that the Saturnian X-ray emission is highly variable — a factor of 3 variability in brightness over one week. These observations also reveal X-rays from Saturn's south polar cap on 20 January (cf. Figure 1, left panel), which are not that evident in our January 26, 2004 (cf. Figure 1, right panel) and earlier CXO observations (Ness et al. 2004b). Surprisingly, from the south polar cap region, no X-rays are detected in the 0.24–0.6 keV band; they are present only in the 0.7–1.4 keV spectral region. This contrasts with Jupiter's aurora, whose emissions are dominated by bands at 0.3–0.4 keV and 0.6–0.7 keV (Branduardi-Raymont et al. 2004; Elsner et al. 2005). The spectral features of X-rays from the polar region are statistically consistent with X-rays from Saturn's disk (see also Ness et al. 2004b). The electron bremsstrahlung mechanism for Saturn's polar x-rays can be excluded on spectral modelling and energetic grounds (Bhardwaj and Gladstone 2000a; Gérard et al. 2004; Ness et al. 2004b). These arguments, along with the fact that the entire south polar cap was sunlit during the Chandra observation, and combined with observations that the Jovian disk X-ray spectrum is harder than its auroral X-ray spectrum (Branduardi-Raymont et al. 2004; Bhardwaj et al. 2004, 2005a), suggest that the X-ray emissions from the south polar cap region on Saturn are unlikely to be auroral in nature; they might instead be an extension of its disk X-ray emission.

Recent Hubble Space Telescope (HST-STIS) observations of Saturn (Clarke et al. 2005) conducted during the January 2004 Cassini approach campaign found that the FUV aurora responds strongly to variations in solar wind dynamic pressure (Crary et al. 2005). This suggests a direct role for the solar wind in powering the FUV aurora on Saturn. Amazingly, our Chandra observations, performed simultaneously with the HST-STIS observations, suggest that the X-ray brightness from the polar region is anti-correlated with the FUV auroral brightness: the X-ray emission was bright on January 20 when the FUV aurora was faint, while the FUV aurora was bright on January 26 when the X-ray aurora was faint. This difference, however, may result from the great difference in the exposure time (Chandra exposures are ~20 times larger than the HST exposures), and the fact that Saturn's aurora in FUV emits photons at a rate more than 1000 times greater than that at X-ray wavelength. However, this may corroborate our spectral findings that suggest that the high-latitude emissions are scattered solar X-rays; and the X-ray aurora is very weak or even absent on Saturn.

To understand the temporal behavior of Saturnian X-rays, we extracted the lightcurve from the region of the disk, minus the south polar cap, and which is not obscured by the rings. Figure 2(a) shows the disk X-ray emission time series for 20 January 2004. Interestingly, it shows a non-uniform variation with time and an occurrence of an 'X-ray flare' on Saturn at around 0730 hr. Coincidentally, the solar X-ray flux measured by GOES satellite at Earth during this time period shows the occurrence of a strong solar X-ray flare at the same time (Figure 2b), once the light travel time difference between Sun-Saturn-Earth and Sun-Earth (134 min) has been taken into account. This class M6.1 flare emanated from sunspot 10540 (NOAA/USAF region), which was located at solar latitude S16 and longitude W12. It lasted for 36 min (started 0734, peaked 0744, and ended at 0810 UT). Since Saturn was near opposition during the Chandra observations (the Sun-Earth-Saturn angle was 159° and the Sun-Saturn-Earth angle was 2°), sunspot 10540 was visible both to Saturn and to GOES. This solar X-ray flare was accompanied by a Hα flare and by radio bursts that included a $4500 \times 10^{-22}$ W/m$^2$/Hz burst at 245 MHz, and caused a sudden ionospheric disturbance at Earth. Analyzing the lightcurve from the south polar cap region, we found that it is almost flat with no evidence of flaring. However, the photons emitted from Saturn's polar cap have a low enough signal to noise ratio that we cannot statistically rule out the presence of flaring.

In addition to the perfect time match between the Saturn and solar X-ray flares, the factor of 5 increase in the Saturn disk X-ray flux during the flare over the average flux-value before the flare is of the same magnitude as the increase in the solar X-ray flux during this time. Moreover, the temporal behavior of the Saturn X-ray lightcurve during the entire observation period on January 20 is similar to that of the solar X-ray flux measured by GOES (cf. Figures 2a, 2b). TIMED/SEE solar X-ray data also indicates a temporal behavior similar to that seen for Saturn X-rays (Figure 2a). Thus, this is the first direct evidence suggesting that Saturn's disk X-ray emission is principally controlled by processes happening on the Sun.

On inspecting the lightcurves of Saturn's disk X-ray emission and solar X-rays for the January 26 observation presented in Figure 3, we find that both lightcurves behave in the same fashion: they are almost featureless and smooth. The X-ray power emitted by the Saturn disk on 26 January is a factor of 3 lower than on 20 January. The GOES X-ray measurement indicates that the solar 1–8 Å flux decreased by a factor of ~7 between 20 and 26 January 2004. This may be consistent with the factor 3 decrease in the power emitted by Saturn in the 6–50 Å band, when the difference in the wavelength bands for GOES and Chandra-ACIS is taken into account, because the variability in the solar X-ray flux is more at shorter wavelengths.

These observations suggest that Saturn's disk X-ray emissions, like those at Jupiter (Bhardwaj et al. 2004, 2005a), are solar X-rays scattered and fluoresced from the planet's upper atmosphere (Maurellis et al. 2000; Cravens et al. 2005). However, not all the incident solar X-rays in the 6–50 Å band are involved: the calculated X-ray albedo of Saturn over this wavelength band is $\sim 7 \times 10^{-4}$, which is similar to that derived by Ness et al. (2004b). The observationally-derived albedo of Saturn is slightly larger than of Jupiter ($\sim 4$–$5 \times 10^{-4}$; cf. Bhardwaj et al. 2004, 2005a). This is consistent with the model of Cravens et al. (2005) and is due to higher He abundance on Jupiter. Intriguingly, the ratio of the number of photons in the 0.24–0.6 keV band to the 0.6–1.2 keV band for the Saturn disk X-rays on January 20 is significantly lower than on January 26: indicating that the solar X-ray flux might have been softer on January 26 than on January 20.

To further demonstrate a relationship between X-rays from Saturn and the solar radiation, we plot in Figure 4 the emitted power in X-rays by Saturn's entire disk for all the observations made so far as a function of solar F10.7 cm fluxes at 1 AU. X-ray fluxes measured by Chandra (our observations and Ness et al. 2004b), XMM-Newton (Ness et al. 2004a) and ROSAT (Ness & Schmidt 2000) X-ray observatories are converted to emitted power using $4 \pi \Delta^2 F$, where $\Delta$ is Earth-Saturn distance and F is the observed energy flux. The F10.7 cm flux has been used as a proxy for the activity of the Sun. The linear relationship between these two parameters adds credence to our conclusion that the Sun directly controls and regulates the X-ray emissions from Saturn.

The current observations, combined with a recent study of Jupiter (Bhardwaj et al. 2005a), demonstrate that the upper atmospheres of the giant planets Saturn and Jupiter act as 'diffuse mirrors' that backscatter solar X-rays. Thus, these planets might be used as potential remote-sensing tools to monitor X-ray flaring on portions of the hemisphere of the Sun facing away from near-Earth space weather satellites. Such a solar flare monitoring instrument does not require high spatial resolution of CXO; it needs to only resolve Saturn. It will also work well for Jupiter resolving its auroral and low-latitude disk X-rays, since Jupiter is about twice the size of Saturn, and auroral X-rays at Jupiter are located at high-latitudes (cf. Gladstone et al. 2002; Elsner et al. 2005). Moreover, unlike XMM-Newton and CXO, such instrument could have a broad spectral band (e.g., 0.1–1.0 keV). Essentially, a modest experiment can work for space weather studies.

**Acknowledgements**

**References**

Bhardwaj, A., & Gladstone, G.R. 2000a, Rev. Geophys. 38, 295.
Bhardwaj, A., & Gladstone, G.R. 2000b, Adv. Space Res. 26, 1551.
Bhardwaj, A., et al. 2002, Eur. Space Agency Spec. Publ. ESA-SP-514, 215.
Bhardwaj, A., et al. 2004, BAAS 36, 1104.
Bhardwaj, A., et al. 2005a, Geophys. Res. Lett. 32, L03S08.
Bhardwaj, A., et al. 2005b, J. Atmos. Sol-Terr. Phys. submitted.
Branduardi-Raymont, G., et al. 2004, A&A, 424, 331.
Clarke, J.T., et al. 2005, Nature 433, 717.
Crary, F.J., et al. 2005, Nature 433, 720.
Cravens, T.E., Waite, J.H., Gombosi, T.I., Lugaz, N., Gladstone, G.R., Mauk, B.H., & MacDowall, R.J. 2003, J. Geophys. Res. 108, 1465.
Cravens, T.E., et al. 2005, in preparation.
Elsner, R.F., et al. 2002, ApJ 572, 1077.
Elsner, R.F., et al. 2005, J. Geophys. Res. 110, A01207.
Gérard, J.-C., Grodent, D., Gustin, J., Saglam, A., Clarke, J.T., & Trauger, J.T. 2004, J. Geophys. Res. 109, A09207.
Gilman, D.A., Hurley, K.C., Seward, F.D., Schnopper, H.W., Sullivan, J.D., & Metzger, A.E. 1986, ApJ 300, 453.
Gladstone, G.R., Waite, J.H., Jr., & Lewis, W.S. 1998, J. Geophys. Res. 103, 20083.
Gladstone, G.R., et al. 2002, Nature 415, 1000.
Krupp, N., et al. 2004, in Jupiter: The Planet, Satellites and Magnetosphere, eds. F. Bagenal, T. Dowling, & W. McKinnon, (Cambridge: Cambridge Univ. Press), 617.
Kurth, W.S. et al. 2005, Nature 433, 722.
Maurellis, A.N., Cravens, T.E., Gladstone, G.R., Waite, J.H., Jr., & Acton, L. 2000, Geophys. Res. Lett. 27, 1339.
McKenzie, D.L., Rugge, H.R. & Charles, P.A. 1982, J. Atmos. Terr. Phys. 44, 499.
Ness, J.-U. & Schmitt, J.H.M.M. 2000, A&A 355, 394.
Ness, J.-U., Schmitt, J.H.M.M. & Robrade, J. 2004a, A&A 414, L49.
Ness, J.-U., Schmitt, J.H.M.M., Wolk, S.J., Dennerl, K. & Burwitz, V. 2004b, A&A 418, 337.
Petrinec, S.M., McKenzie, D.L., Imhof, W.L., Mobilia, J. & Chenette, D. 2000, J. Atmos. Sol-Terr. Phys. 62, 875.
Russell, C.T. 2001, Planet. Space Sci. 49, 1005.
Snowden, S.L. & Freyberg, M.J. 1993, ApJ 404, 403.
Stadsnes, J., Aarsnes, K. & Bjordal, J. 1997, Adv. Space Res. 20, 1043.
Stallard, T.S., et al. 2004, Icarus 167, 204.
Waite, J.H., Jr., et al. 1997, Science, 276, 104.
Waite, J.H., Jr., & Lummerzheim, D. 2002, in Atmospheres in the Solar System: Comparative Aeronomy, eds. M. Mendillo, et al., AGU Geophys. Monograph 130, (AGU, Washington, DC), 115.

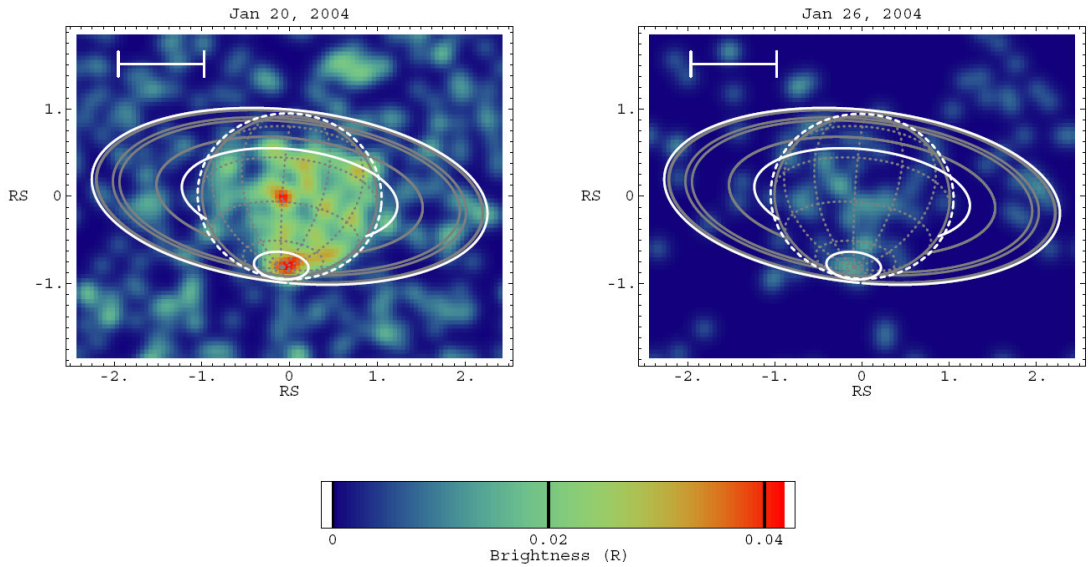

**Figure 1.** Chandra ACIS X-ray 0.24–2.0 keV images of Saturn on 20 and 26 January 2004. Each continuous observation lasted for about one full Saturn rotation. The false color images, with brightness in Rayleighs (R), show X-ray photons as seen in a frame moving across the sky with Saturn, smoothed with a two dimensional gaussian with $\sigma = 0.984''$ (twice the ACIS pixel width). The horizontal and vertical axes are in units of Saturn's equatorial radius. The white scale bar in the upper left of each panel represents $10''$. The superimposed graticule shows latitude and longitude lines at intervals of $30°$. The solid grey lines are the outlines of the planet and rings, with the outer and inner edges of the ring system shown in white. The dotted white line defines the region within which events were accepted as part of Saturn's disk unless obscured by the rings. The white oval around the south pole defines the polar cap region. The conversion to Rayleighs was performed using a value of the average effective area of 447 cm$^2$, which was the average over the nominal energies of all events from Saturn. The two images taken a week apart, and shown on the same color scale, indicate substantial variability in Saturn's X-ray emission.

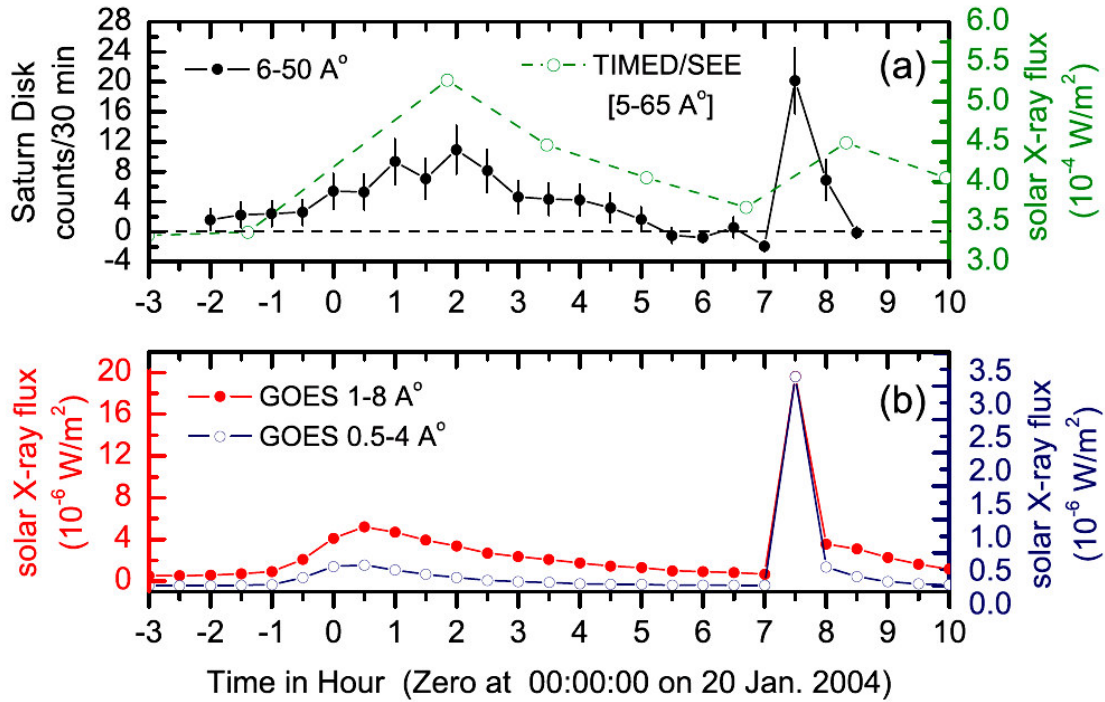

**Figure 2.** Lightcurve of X-rays from Saturn and the Sun on 20 January 2004. All data are binned in 30-minute increments, except for the TIMED/SEE data which are 3-minute observation-averaged fluxes obtained every orbit (~12 measurements per day). **(a)** Background-subtracted low-latitude (non-auroral) Saturn disk X-rays (0.24–2.0 keV) observed by Chandra-ACIS, plotted in black (after shifting by -2.236 hr to account for the light travel time difference between Sun-Saturn-Earth and Sun-Earth). The solar 0.2–2.5 keV fluxes measured by TIMED/SEE are open green circles and are joined by the green dashed line for visualization purpose. **(b)** The solar X-ray flux in the 1.6–12.4 keV and 3.1–24.8 keV bands measured by the Earth-orbiting GOES-12 satellite. A sharp peak in the lightcurve of Saturn disk X-ray flux — an X-ray flare — is observed at about 7.5 hr, which corresponds in time and magnitude with an X-ray solar flare. In addition, the temporal variation in Saturn's disk X-ray flux during the time period prior to the flare is similar to that seen in the solar X-ray flux.

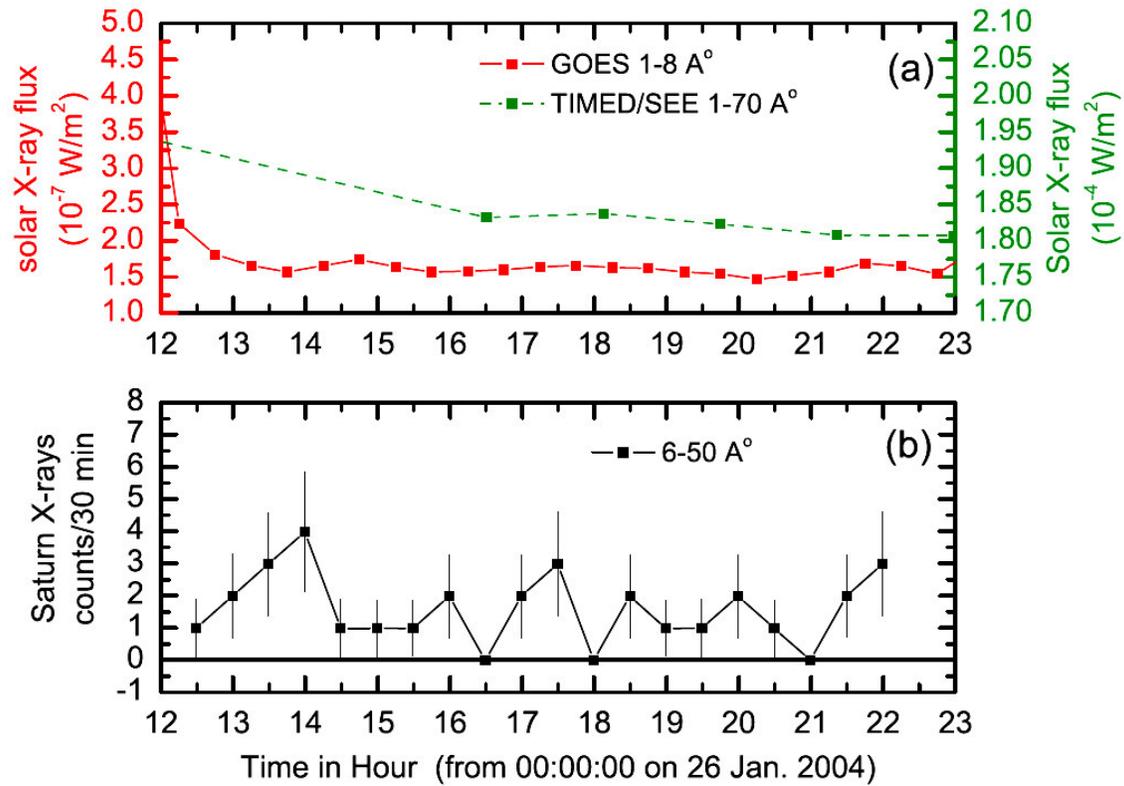

**Figure 3.** Lightcurve of background subtracted X-rays from the low-latitude (non-auroral) disk of Saturn observed by Chandra ACIS along with those of the solar X-rays on 26 January 2004. All data are 30-min binned, except TIMED/SEE data which are 3-min observation-averaged fluxes obtained every orbit (~12 measurements per day). **(a)** Solar X-ray fluxes in 1.6-12.4 keV band from the GOES 12 and 0.2-2.5 keV band from the TIMED/SEE (connected by dashed line for visualization purpose). **(b)** Saturn low-latitude disk 0.24-2.0 keV X-ray emission, plotted after shifting by -2.241 hr to account for light travel time delay between Sun-Saturn-Earth and Sun-Earth. Both the Saturn and solar X-ray flux show a similar variation with time with a flat-type profile.

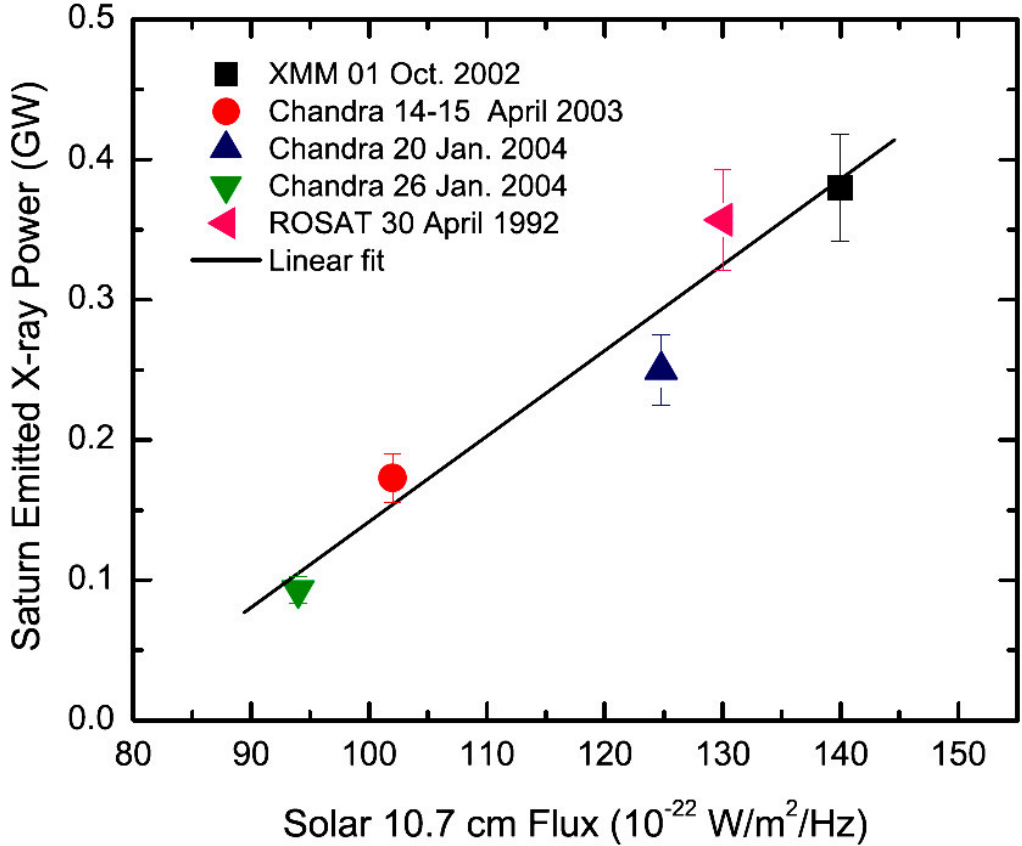

**Figure 4.** The X-ray power emitted from Saturn's disk plotted against the value of the solar 10.7 cm flux on the day of observation. X-ray fluxes measured by Chandra, XMM-Newton and ROSAT X-ray observatories are converted to emitted power (see text). A uniform 10% error bar is shown for all observations. The energy flux for Chandra and XMM-Newton observations are for a similar energy range of ~0.2–2.0 keV, while for the ROSAT observation it is 0.1–0.55 keV. The solid black line shows a linear fit to the emitted power. The correlation in the X-ray power emitted from Saturn's disk with the solar 10.7 cm flux suggests that the two parameters are closely related, and implies that X-ray emission from the disk is primarily controlled by solar radiation.